# HAPTIC FEEDBACK SYSTEMS IN MEDICAL EDUCATION

**Felix G. HAMZA-LUP[*], Dorin M. POPOVICI[**], Crenguta M. BOGDAN[***]**

*Abstract:* This paper brings into discussion some of the most relevant technological challenges involving haptic systems in medical education. One of these challenges is choosing the suitable haptic hardware, API or framework for developing a visuo-haptic eLearning system. The decision is based on several criteria such as the multimodal resources needed by the software system, compatibility with haptic devices and the dynamic configuration of the scene. Another challenge is related to the software system reactivity in conjunction with the user's actions. The immediate haptic feedback from the virtual models, together with the synchronization of the rendered haptic and visual cues seen by the users are essential for enhancing the user's learning ability. Visuo-haptic simulation facilitates accurate training scenarios of medical protocols and surgical processes.

## I. Introduction

**H**aptics facilitate the introduction of tactile sensation in computer applications, enabling users to receive feedback they can feel in addition to other cues like auditory and visual. Multimodal environments where visual, auditory and haptic stimuli are present have an increased potential to convey information more efficiently since the user manipulates and experiences the environment through multiple sensory channels.

[*] *Department of Mathematics and Computer Science, Ovidius University of Constanta, 124 Mamaia Blvd., 900537, Constanta, Romania; Department of Computer Science and Information Technology, Armstrong Atlantic State University, Savannah 31419, Georgia, USA.*
[**] *Department of Mathematics and Computer Science, Ovidius University of Constanta, 124 Mamaia Blvd., 900537, Constanta, Romania.*
[***] *Department of Mathematics and Computer Science, Ovidius University of Constanta,124 Mamaia Blvd., 900537, Constanta, Romania.*

The availability of haptic systems enables the augmentation of traditional instruction with interactive interfaces offering enhanced motivation and intellectual stimulation. Although the haptic devices have not made large inroads into medical education yet, the potential for revolutionary change exists due to the recent availability and affordability of both hardware and software components.

## II. Haptics in Medical Education

The potential of haptic interfaces was initially confirmed in various medical training applications like cardiology [1], prostate cancer diagnosis [2], injection [3] and lumbar punctures procedures [4], surgery [5,6] and angioplasty interventions [7], rhinoscopy and bronchoscopy procedures [8], palpatory diagnosis [9], but also in orthopaedic drilling [10] and bone surgery procedures [11]. In [3], the use of haptic devices is investigated in combination with speech input and output as physical interaction triggers supplementary emotions at the virtual patient level.

Dentistry procedures (as implants [12] or dental preparations [13]) are usually trained using mixed realities (MR) in combination with haptic feedback. The system proposed by [14,15] allows students to practice surgery using the correct postures as in the actual environment by mixing 3D tooth and tool models upon the real-world view and displaying the result through a video see-through head mounted display (HMD). Their system addresses different student skills levels by incorporating an adaptable kinematic feedback and hand guidance using haptic devices, by comparison with an expert's gestures.

Guidance is also used in palpation-based (i.e. palpatory) systems as the Virtual Haptic Back project at Ohio University that implements a haptic playback system using the PHANToM haptic interface. The haptic interface is used for students training, to improve virtual palpatory diagnosis by allowing the user to follow and feel an expert's motions prior to performing their own palpatory tasks on the patient [16].

Virtual Veins is a virtual reality (VR) training simulator allowing healthcare practitioners to acquire, develop and maintain the skills required to perform venipuncture in a range of realistic scenarios within a safe controlled environment [7]. After a practice or test session, the user can review their own performance in an online report. The report contains user and session details, an event log of the session (i.e. skin/vein penetrations/retractions), a screenshot of the skin penetration as well as metrics including skin insertion angle, skin retraction angle, bevel angle, vein insertion angle, vein diameter, penetration length and penetration depth.

Rehabilitation is another field closely related with medicine where haptic feedback can be employed. In [17] a haptic system for hand rehabilitation combines robotics and interactive VR to facilitate repetitive performance of task specific exercises for patients recovering from neurological motor deficits. Many other haptic-based rehabilitation systems are under research and development.

The current accessibility of haptic technology together with the intuitiveness of using it represents enough arguments for applying haptic feedback in education at all levels, starting from elementary schools [18]. Dedicated setups have been proposed for teaching elementary school students simple-machine concepts [19] as well as complex ones [20]. Dynamic systems have been proposed [21] as well as modelling virtual proteins [22, 23] and experiencing multimodal educational virusology-oriented content [24]. Haptic devices are used as tools for sculpturing [25] and even for developing writing skills [26]. All these haptic-oriented solutions rely on the fact that haptics simulates and renders realistically the interaction forces that occur when the user encounters real objects. Haptic feedback enables the user to become more engaged in the proposed physical experience, not only by traditional means (mental, visual, and auditory) but also from a tactile perspective, improving the immersive component of the VR experience in the educational process because of the orthogonal perspectives given by the sensorial data [27].

### III. Haptic Devices

The haptic devices currently available on the market apply relatively small forces on the user (usually on the user's hands and/or fingers) through a complex system of servomotors and mechanical links. There are numerous haptic devices on the market, and their price has decreased significantly over the past decade due to mass production. Among the most popular are Sensable's PHANToM® Omni™ (Figure 1) and Desktop™ [28] devices that can apply forces through a mechanical joint in the shape of a stylus. As recent as 2007, Novint, a company founded by the researchers of Sandia National Laboratory, marketed the very first commercial haptic device. Falcon Novint [29] (Figure 1) has been released on the market at a very low price in conjunction with computer games in the USA, Asia and Australia. Novint licensed key portions of the technology used in Falcons from Force Dimension [30], a leading Swiss developer of high-end haptic devices like the Omega family.

A novel approach to implementing haptic feedback is through magnetic forces. Magnetic levitation haptic devices allow users to receive force-feedback by manipulating a handle that levitates within a magnetic field. Users can translate and rotate the handle while feeling forces and torques from the virtual environment.

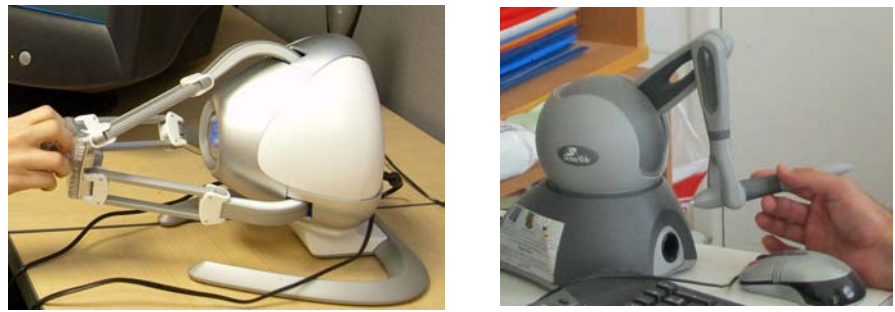

**Figure 1.** Falcon Novint (left), Sensable Omni (right)

Compared with traditional haptic devices that use motors, linkages, gears, belts, and bearings, magnetic levitation uses a direct electro-dynamic connection to the handle manipulated by the user. Some of the advantages of this approach include no static friction, no mechanical backlash, high position resolution, simulation of a wide range of stiffness values, as well as mechanical simplicity. Magnetic haptics is considered in relation to surgical training systems [31]. The first commercial integration of a magnetic levitation haptic device is the Maglev 200™ Haptic Interface developed by ButterflyHaptics™ [32].

The most important characteristics (i.e., performance measures) [33] common to all haptic devices, include degrees-of-freedom, work-volume, position resolution, continuous force, maximum force/torque, maximum stiffness, update rate and inertia. Find a balance among all these factors is a real challenge.

Multiple problems arise in haptic applications interacting with deformable objects. For example, costly computation time, numerical instability in the integration of the body dynamics, and time delays may occur. Lengthy computations are an issue in haptic systems because of the high update rates (approx. 1 KHz) to obtain realistic force feedback. The update rates of the visual component (i.e., graphic rendering) of the physical objects being simulated is of the order of 30 to 60 frames per second (Hz).

This difference in the simulation rates can cause an oscillatory behaviour in the haptic device that can become highly unstable. Some of these problems can be alleviated with the use of magnetic levitation devices [32]; however, the development of applications in the area is in early research stages.

### IV. Haptic APIs and Frameworks

In one of our research studies [34], we synthesized information regarding existing haptic APIs, presenting the now "extinct" ReachIn commercial API, and other active open-source APIs like SOFA, CHAI3D, H3D, X3D, GiPSi and OpenHaptics.

ReachIn [35] was one of the first haptic development platforms that enabled the development of sophisticated haptic 3D applications in the user's programming language of choice, such as C++, Python, or VRML (Virtual Reality Modeling Language). This API was one of the first commercial ones that involve haptics. It's structure allows the development of multimodal interfaces and synchronizes haptic, graphic, audio or non-haptic devices. The company was restructured a few years ago and the ReachIn API is no longer active. However, the experiences learned from this API are quite important for the haptic research community.

SOFA (Simulation Open-Framework Architecture) [36] is an international, multi-institution, collaborative initiative, aimed at developing a flexible and open source framework for interactive simulations. Using a scene graph structure, SOFA provides several views in modeling 3D objects: a dynamic view that include masses and constitutive laws for the objects, a collision view that use simplified 3D models of the objects in collision computation, and a visual view that uses a complex 3D graphical representation. SOFA assures the scene consistency between these models by using mapping modules. Moreover, SOFA implements complex real-time algorithms that use multiple representations of the simulated objects in each view.

Computer Haptics and Active Interfaces - CHAI3D [37] is an open-source designed to facilitate the development of 3D modelling applications augmented with haptic rendering. It supports several commercial haptic interfaces such asServo2Go and Sensoray 626 I/O board, IEEE1394 interface.

CHAI3D provides an easy solution to interface any haptic device with a specific computer-based application. CHAI3D framework allows extensions using modules for ODE [38] and dynamic engines that simulate rigid and deformable objects in real-time. Moreover CHAI3D enables the development of new classes, in order to integrate new haptic and visual rendering algorithms as well as drivers for new devices.

A popular open-source platform, H3D [39] is dedicated to haptic modelling that combines the OpenGL and X3D standards together with haptic rendering in a single scene graph that mixes haptic and graphic components. H3D is independent of haptic device multi-platform that allows audio and 3D stereoscopic device integration. H3D is conceived to support rapid prototyping. Combining X3D, C++ and the Python scripting language, H3D improves the speed of execution, when performance is critical, as well as high speed of development, when rapid prototyping is required.

General Physical Simulation Interface (GiPSi) [40] is an open-source framework that presents a flexible architecture, developed to simulate surgical procedures at organ level. The architecture interconnects computational and data models, developed by different research teams, quantitative validation of biological simulations together with software modules interconnections.

OpenHaptics toolkit [41] developed by SenseAble, includes the QuickHaptics interface, the haptic device interface, the haptic library interface, together with tools and drivers for Phantom® devices. Comprehensive documentation and a programmer's guide accompany the toolkit.

### V. Our Experience with Haptic Systems Design and Development

The CeRVA team [42] (Research in virtual and augmented reality) from the Faculty of Mathematics and Computer Science, Ovidius University Constanta started to experience the use of haptic devices in 2008, in the framework of Virdent project (PNII 12083/2008). The Virdent project (Figure 2) was oriented towards the preparation simulation in fixed dental prosthesis. It offered a non-invasive, feasible, simulation providing the necessary feedback for learning and offering students the capacity to recover from erroneous situations. The simulator included recording and real-time evolution of the student evaluation by the teacher [15].

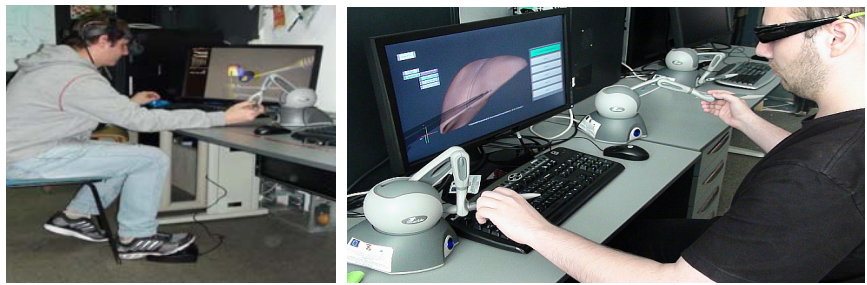

**Figure 2.** Working sessions with Virdent (left) and HapticMed (right)

In 2010 under Dr. Felix Hamza-Lup leadership, we started the HapticMed project (POSCCE O.2.1.2/2009), the first 3D visual and haptic simulator for liver diagnostic through palpation in Romania (see fig. 2). This custom-built simulator has enabled development of new expertise in haptic system development and integration for Romanian computer science and engineering students [43].

As opposed to commercial simulators for laparoscopic procedures, the proposed simulator cost is a fraction of the cost of any existing commercial simulator and was developed mainly with open source software. Moreover, the simulator is flexible and reconfigurable for different procedures. The results obtained so far point to direct applications in the medical industry and practice. The simulator can improve medical training thus helping save human lives.

We are in the process of assessing the simulator through an ample assessment effort in collaboration with surgeons and medical residents from the Regional Hospital of Constanta.

## VI. Conclusions

The potential of haptic interfaces in support of practice-based learning is established worldwide. Even if the haptic hardware is more affordable, the development of haptic-based simulations is hampered by the lack of existing software adaptability and extensibility as well as by the lack of student training in real-time systems and software development in this field. We proved that with appropriate expert involvement in the development process, we could develop haptic-based prototypes as well as create and improve competencies in the field. We believe that haptic-based simulators will become widespread in the near future and underline the need to develop expertise in this field in Romania

**Acknowledgements**

This study was supported under the ANCS Grant "HapticMed – Using haptic interfaces in medical applications", no. 128/02.06.2010, ID/SMIS 567/12271, POSCCE O.2.1.2 / 2009. We would like to thank Aurelian Nicola, Elena Bautu, Andrei Vasilateanu, Cristian Corleanca as well as the other members of the CeRVA team from Ovidius University of Constanta, Romania for their support.